\documentstyle[twocolumn,aps,prl]{revtex}

\begin{document}

\draft
\title{Critical and supercritical dynamics of quasiperiodic systems}
\author{ Jukka A. Ketoja}
\address{Department of Mathematics, University of Helsinki, P. O. Box 4,
FIN-00014 Helsinki, Finland}
\author{Indubala I. Satija\cite{email}}
\address{
 Department of Physics and
 Institute of Computational Sciences and Informatics,\\
 George Mason University,
 Fairfax, VA 22030, USA}
\maketitle
\begin{abstract}
The almost periodic eigenvalue problem described by the
Harper equation is connected to 
other classes of quasiperiodic behaviour:
the dissipative dynamics on critical invariant tori and
quasiperiodically driven maps. 
Firstly, the strong coupling limit of the supercritical
Harper equation and the strong dissipation limit of the critical
standard map play equivalent role in the renormalization analysis
of the self-similar fluctuations of localized eigenfunctions
and the universal slope of the projected map
on the invariant circle.
Secondly, we use a simple transformation to relate the Harper equation to
a quasiperiodically forced one-dimensional map. In this case, the localized 
eigenstates of the supercritical Harper equation 
correspond to strange but nonchaotic attractors of the driven map.
Furthermore, the existence of localization in the
eigenvalue problem is associated with the appearance of
homoclinic points in the corresponding map. 
\end{abstract}

\narrowtext

\section{Introduction}

Systems with two competing periodicities have been in the forefront
of research in nonlinear dynamics as well as in condensed matter
physics. Perhaps the most well-known paradigm in the study of
quasiperiodic dynamics  in autonomous dissipative and Hamiltonian
systems is the standard map with and without dissipation\cite{MacKay,dstd}.
A related problem in the condensed matter physics is
the Schr\"odinger
equation with quasiperiodic potential \cite{Mathieu}. A simple discrete
prototype in this field is the Harper equation \cite{Harper} or
the almost Mathieu equation \cite{Mathieu}.
Invariant circles in the standard map correspond to extended
states of the Harper
equation while the breakup of an invariant circle is
analogous to the localization transition in the Harper equation.
Both transitions have been studied by the
renormalization methods\cite{MacKay,Kadanoff,Rand,Ostlund}.

We have applied the decimation methodology, developed earlier
to show the existence of
"slope" universality in the dissipative, critical standard map\cite{K}, to
a variety of eigenvalue problems
which can be written in the linear-difference form \cite{KS}. Very recently,
the renormalization analysis was extended to the Harper
equation beyond criticality \cite{KSloc}.
These studies have shown the existence of a unique strong coupling
fixed point which describes the fluctuations of localized eigenstates
in the whole of supercritical region.
In the first part of this paper, we show that this strong coupling
fixed point of the
Harper equation is related to the strong dissipation limit of the critical
standard map. The renormalization analysis for the critical
standard map is rewritten in the form which makes the 
correspondence clear.

The second part of this paper is to relate the results of the
supercritical Harper equation to the dynamics of quasiperiodically
 forced dissipative maps. In parallel to the study of
quasiperiodic dynamics in autonomous dissipative and Hamiltonian
systems there have been investigations on the dissipative dynamics
of quasiperiodically forced systems. Although the presence
of such forcing causes some usual orbits like the periodic ones
to be absent in these systems, there are other interesting features which
have not been found in autonomous systems. Perhaps the most striking
phenomenon is the appearance of strange nonchaotic attractors
(SNA) \cite{GOPY}. 
These are geometrically strange attracting sets for which all the
Lyapunov exponents (except the one related to the forcing) are
negative. Inspite of a great deal of research in systems exhibiting 
SNA \cite{forced,KPF}, there are still many open questions:
the condition for the existence of SNA as well as a general
renormalization theory on their appearance are not fully understood.

Bondeson et al. \cite{Bon} pointed out an interesting connection
between a quasiperiodically forced oscillator and the Schr\"odinger
equation with quasiperiodic potential. Using the
substitution introduced by Pr\"ufer \cite{Pru}
already seventy years ago,
the two classes of systems can be mapped onto one another the energy
eigenvalue of the Schr\"odinger equation appearing as a parameter for
the oscillator. The transition from an invariant circle
to SNA in the classical problem is related to the transition from
extended to localized states in the quantum problem.

In this paper, we map the discrete Schr\"odinger equation, namely the
Harper equation, to a quasiperiodically driven one-dimensional map
so that the eigenfunction corresponds to an attractor of the map.
The simplicity
of the discrete system compared to the continuum model offers the
possibility of obtaining further insight into the existence and
 characterization of
SNA. By focusing on the strong coupling limit, we are able to show the
existence of SNA in quasiperiodically driven dissipative maps. 
Furthermore, this approach 
also helps in understanding
some of the open questions related to the significance of the critical phase
in the Harper equation.\cite{Ostlund,KSloc}
Our focus is somewhat similar to the one of Bondeson et al. \cite{Bon}
in that we are interested in various transitions and their effects on
the renormalization dynamics or the dynamics of the forced map.

In Section II we briefly review the renormalization approach as applied
to the Harper equation and the dissipative standard map.
In Section III, we show how these two problems are related.
In Section IV, we discuss the dynamics of the one-dimensional
map associated with quasiperiodic eigenvalue problems and for
the first time
show explicitly the connection
between localization and strange
nonchaotic attractors conjectured earlier by Bondeson et al. \cite{Bon}.
In Section V, we discuss our results from a general perspective. As a 
comparison with previous two quasiperiodic systems, we describe the results
with completely different properties for
the phonon equation of the Frenkel-Kontorova model although
this eigenvalue problem
is formally very close to the Harper equation.

\section{Renormalization approach to quasiperiodic systems: 
the Harper equation and the dissipative standard map}

We have developed a decimation approach to study quasiperiodic systems
with two incommensurate frequencies which can be written in the following
linear-difference form:\cite{K,KS,KSloc}
\begin{equation}
a(i)\psi_{i+1}+b(i)\psi_{i-1}+c(i)\psi_i = 0 . \label{TBM}
\end{equation}
Here, $a$, $b$, and $c$ are real or complex functions of 
$i\sigma$ where $\sigma$ is an irrational number equal to the ratio 
of two frequencies. Although the decimation method can be implemented
 for any irrational $\sigma$,
here we consider the simplest case where $\sigma$ is given by
the inverse golden ratio $\sigma =(\sqrt 5 -1)/2$.
Our formulation can be generalized also to
the case where $\psi_i$ is a multi-component
vector and $a$, $b$, and $c$ are matrices.

In the decimation scheme, it is appropriate to decimate out
all sites except those labelled by the Fibonacci
numbers $F_n$ (which are the best rational approximants of
the golden ratio). At the $n^{th}$ decimation level $(n=2,3,...)$,
the linear-difference equation is expressed in the form
\begin{equation}
f_n(i) \psi(i+F_{n+1})=\psi(i+F_n) + e_n(i) \psi(i). \label{Dec}
\end{equation}
The additive property $F_{n+1} =F_n + F_{n-1}$
of the Fibonacci numbers provides exact recursion
relations for the decimation functions $e_n$ and $f_n$:
\begin{eqnarray}
e_{n+1} (i)= - {A e_n (i) \over 1+Af_n (i)} \\
f_{n+1} (i)= {f_{n-1} (i+F_n) f_n(i+F_n)\over 1+Af_n(i)} \\
A = e_{n-1} (i+F_n) + f_{n-1} (i+F_n)e_n(i+F_n). \nonumber
\end{eqnarray}

The decimation functions $e_n$ and $f_n$ define a renormalization flow
which converges asymptotically on an attractor. In the case when the attractor
is a limit cycle of period $p$, it reflects the translational invariance
of the function $\psi$ in the Fibonacci space and also captures the 
self-similarity underlying it at all scales.
The decimation functions $e_n$ and $f_n$
determine the universal scaling ratios
\begin{equation}
\zeta_j = \lim_{n \rightarrow \infty} |\psi(F_{pn+j})/\psi(0)|;\;\;j=0,...,p-1.
\end{equation}
\hfill\break

The two quasiperiodic systems that we discuss here are the Harper equation
\begin{equation}
\psi_{i+1} + \psi_{i-1} + 2\lambda \cos[2\pi (i\sigma +\phi)] \psi_i =
E\psi_i  \label{Harp}
\end{equation}
and the dissipative standard map
\begin{equation}
x_{i+1} = x_i + \Omega +b(x_i -x_{i-1} ) -{k\over 2\pi} \sin(2\pi x_i) .
\end{equation}
In the Harper equation, $E$ is the
eigenvalue corresponding to the eigenfunction $\psi_i$. For irrational
$\sigma$, the equation describes a particle in a periodic potential
incommensurate with the periodicity of the lattice.
On the other hand, the standard map with $0 \leq b < 1$, and for
certain values of $\Omega$, describes quasiperiodic dynamics
on an invariant circle.  This equation can be written in the 
linear-difference form 
by considering the equation for the derivatives
$\xi_i =\partial x_i /\partial x_0$ describing the slope of the $i$
times  iterated reduced circle map $x_{k+1} =h(x_k )$ at $x_0$:
\begin{equation} 
\xi_{i+1} + b\xi_{i-1}
- [1+b -k\cos(2\pi x_i )]\xi_i =0 .\label{slopes}
\end{equation}

In ref. \cite{K}, a somewhat similar formulation of the
decimation equation (\ref{Dec}) and the recursion relations
was used to derive the following result  along the 
critical line
$k=k_c (b,\Omega )$ for the breakup of the golden
invariant circle in the standard map:
$\xi_{F_n }b^{-F_n} \to \zeta_S $ as $n\to \infty$, where
$\zeta_S=0.435625...$ and the slope was calculated at the point $x_0$
which lead to the smallest value. In fact, the value
of $\zeta$ was calculated in the limit $b\to 0$ and its
universality, termed as the ``slope universality'',
for finite $b$ was not understood although suggested by the numerical results.
This result, which has not been yet
rigorously proven, would have an interesting mathematical implication:
under renormalization,
the critical renormalization fixed point would attract not only circle
maps with zero slope but also those with finite slope.
The latter would correspond to critical reduced circle maps of
higher dimensional systems with finite Jacobians. They would give a new 
dimension for the stable manifold of the critical renormalization
fixed point.

Later \cite{KS}, the same decimation procedure was applied to the 
critical Harper equation with $\lambda =1$ for fixed $E$ which
was usually chosen either at the band center or at the band edges. Here
the $p$ universal numbers $\zeta$ 
described the amplitude of the wave function at
points which were spaced $p$ (or any multiple of $p$)
Fibonacci sites apart from the central peak.
More recently \cite{KSloc},
the modified decimation procedure explained in above
was used to describe universal
features in the Harper equation $beyond$ criticality.\cite{KSloc}
Here, the fluctuations of localized eigenfunctions
were shown to be universal.

In the next section, we discuss both problems using this
modified decimation approach and show that
the {\it supercritical} Harper equation with $\lambda > 1$ and the
{\it critical} dissipative
standard map with $b < 1$ are closely related.
In particular, this analysis clarifies the universality
of $\zeta_S$ for the full range of the dissipation parameter $b$.

\section{Analogy between the supercritical Harper equation and
the critical dissipative standard map}

In this section, we explain the common principle behind
the slope universality of the critical dissipative standard map
\cite{K} and the supercritical universality of the Harper equation
\cite{KSloc}.
In both cases, the universality can be generalized also to other than 
the Fibonacci sites but for the present purpose it suffices
to consider this restricted universality.

The Harper equation in the supercritical regime ($\lambda > 1$)
is characterized by the positive inverse localization length
$\gamma =ln(\lambda)$ for the wave function $\psi_i$ \cite{Aubry}.
The fluctuations of the localized eigenfunction are obtained by
factorizing the exponentially decaying part out.
The wave function $\psi_i$ is written as
\begin{equation}
\psi_i = e^{ -\gamma i} \eta_i .
\end{equation}
 The resulting equation for the
fluctuations is given by the linear-difference form
\begin{equation} \label{fluctu}
{1 \over \lambda} \eta_{i+1} + \lambda \eta_{i-1} + 2 \lambda
\cos[2 \pi (i\sigma +\phi)]
\eta_i = E \eta_i .
\end{equation}
A correspondent equation for the standard map is obtained by 
factorizing out the exponentially decaying effective Jacobian
from the slope. In other words, with $\xi_i = b^i \eta_i$
we obtain
\begin{equation} 
b \eta_{i+1} +\eta_{i-1} - [1+b -k\cos(2\pi x_i )]\eta_i =0 .\label{slopefluct}
\end{equation}

It is interesting to note that the above equations have well-defined
limits when $\lambda \rightarrow \infty$ or $b\to 0$,
respectively. Moreover, these limits 
resemble one another.
As the latter corresponds to the circle map,
for which the existence of a renormalization fixed point has been known
for a long time \cite{Kadanoff,Rand}, the analogy suggests 
the existence of a similar
renormalization fixed point also for the fluctuations in the Harper equation
above criticality.
In fact, applying the decimation tools to Eq. (\ref{fluctu}),
the fluctuations of 
states with eigenvalue either at the band center or at the band edges
were found to be self-similar characterized by periodic
attractors of the renormalization flow \cite{KSloc}.
In particular, for the states at the band edges with
$\phi=1/4$, $|\eta_{F_n } /\eta_0| \to \zeta_H =0.172586410945... $
 as $n\to \infty$, independently of 
the value of $\lambda > 1$. This universality can also
be interpreted in an alternative way: 
the finite-size inverse localization length $\bar{\gamma}$,
defined by $\psi_i = e^{-\bar{\gamma} i} $,
varies around the asymptotic value $\gamma$ in a universal way
given by the equation
\begin{equation}
\gamma - \bar{\gamma} = \log |\eta_{F_n}|/F_n .
\end{equation}
This is a rather intriguing result
as the inverse localization length does depend upon $\lambda$.

In order to understand the origin of the parameter-independent
universality in the Harper equation and the dissipative standard map,
we write the equation for $\eta_i$ in both cases in the form
\begin{equation}
f_2 (i) \eta(i+2) = \eta(i+1) + e_2 (i) \eta(i) 
\end{equation} 
which gives the first step of a decimation approach of generating
equations of the form
\begin{equation}
f_n (i) \eta (i+F_{n+1}) = \eta(i+F_n ) + e_n (i) \eta(i) .
\end{equation}
The power of this method lies in the fact that in iterating
the recursion relations the decimation function $f_n$ asymptotically
``renormalizes'' to zero. This is a numerical observation but also
suggested by the appearance of the constant $C=1 /\lambda<1$ (for
the Harper equation) or $C=b<1$ (for the standard map) in front
of the function $f_2$. Because one can take the trivial conditions
$f_1 \equiv 1$ and $e_1 \equiv 0$, in $f_n$ there appears
a multiplying constant $C^{F_{n-1}} $ which tends to zero
as $n \to \infty$. In other words, asymptotically the recursion
relation for the decimation function $e_n$ simplifies into  
\begin{equation}
e_{n+1} (i) = - e_{n-1} (i+F_n ) e_n (i)  
\end{equation}
which is the form in the strong coupling limit $\lambda \to \infty$
of the Harper equation or the infinite dissipation limit $b\to 0$
of the standard map. We can therefore expect the renormalization
fixed points in these limits to describe the scaling properties
of $\eta(i)$ for the whole $\lambda$ or $b$ range.
One should also note that because $f_n$ vanishes asymptotically,
the scaling ratio $\zeta$ is given simply by the asymptotic value of
$e_n$ at the lattice site $i=0$.

In order to solve for the fixed points, the discrete lattice index
$i$ has to be replaced by a continuous variable. For the standard
map, we can take this to be simply the original variable $x$ so that
the recursion relation becomes
\begin{equation}
e_{n+1} (x) = - e_{n-1} (h^{F_n} (x)) e_n (x)  .
\end{equation}
This equation appeared also in the previous work \cite{K}.
However, because there a variant of the decimation
method was applied directly to Eq. (\ref{slopes}) and not to Eq.
(\ref{slopefluct}), 
one was left with another non-trivial decimation function for
positive $b$. This hindered recognizing the role played
by this equation for finite Jacobians.
Because of the universal scaling properties of the reduced
circle map $h$, $e_n (x_0)$ approaches a universal limit which can be
obtained in terms of the universal fixed point
of the circle map renormalization without solving $e(x)$
for arbitrary $x$ \cite{K}.
However, one has to be careful in the analysis because
effectively it means calculating the limit of
$\xi_{F_n} / b^{F_n} $ taking both the numerator and
the denominator to zero (and $n$ to infinity).

In the case of the Harper equation, the continuous variable 
for $e_n$ is
obtained from the fractional part $\{ i\sigma \}$ of $i\sigma$,
$x=(-\sigma)^n \{ i\sigma \} $, and the resulting
fixed point equation
\begin{equation}
e^* (x) = - e^* (\sigma^2 x+\sigma) e^* (-\sigma x)  \label{HarpStrong}
\end{equation}
 is solved by standard expansion methods \cite{KSloc}.
We show this universal function in Fig. 1 which illustrates the
nontriviality of the strong coupling fixed point as compared to
the trivial weak coupling fixed point \cite{Ostlund,KS}
of the Harper equation.
As seen in the plot, the function is smooth and
finite $almost$ everywhere.

It is interesting that the above fixed point equation (\ref{HarpStrong})
was independently found
by Kuznetsov et al. \cite{KPF} in studying the birth of a strange nonchaotic
attractor in a quasiperiodically forced map.
In the next section, by using a 
discrete version of the Pr\"ufer transformation, we relate the
Harper equation directly to such a map.

\section{Localization and strange nonchaotic attractors}

The Pr\"ufer substitution \cite{Pru} that was used to
map the continuous Schr\"odinger equation
to overdamped, quasiperiodically driven oscillator has motivated us
to consider a similar transformation also in the discrete case.
The main idea underlying the continuous transformation is to define
new variables  $\rho$ and $\alpha$ in terms of the
wave function $\psi=\rho \cos (\alpha)$ and its derivative
$\psi' =\rho \sin (\alpha)$ so that one obtains a first order
equation for $\alpha$ where $\rho$ does not appear and where
the potential of the quantum problem appears as a driving term.
An analogous idea in the case of the discrete Harper equation would
be to set e.g. $\psi_{i-1} = \rho_i \cos(\alpha_i )$,
$\psi_i  = \rho_i \sin(\alpha_i )$. In fact, the Harper equation
gives trivially a very simple equation for $x_i = \cot (\alpha_i ) =
\psi_{i-1} / \psi_i $:
\begin{equation}
x_{i+1} = {-1 \over x_i - E + 2\lambda \cos[2\pi(i\sigma +\phi)]} .\label{map}
\end{equation}
It is clear that higher iterations of this map generate continuous
fraction expansions in terms of subsequent 
potential terms $V(i)=\lambda \cos[2\pi(i\sigma +\phi)]$.
In the following, we study this one-dimensional map and
verify some of the conjectures of Bondeson et al. \cite{Bon}
in the discrete context. 
Although the transformation above is extremely simple, it
provides a useful tool to study various eigenvalue problems.

Numerical study of the map (\ref{map}) shows that the
extended phase of the Harper equation
corresponds to invariant curves for the
map. There are infinity of such
curves depending upon the initial condition.
The transition to localization in the Harper equation
corresponds to transition from invariant curves to an attractor.
Fig. 2 shows the numerically obtained
 Lyapunov exponent of the map as a function
of the parameter $E$. As seen from the figures,
the eigenenergies are singled
out in this plot by giving rise to Lyapunov exponents larger than those
of forbidden values of $E$. Furthermore, 
the Lyapunov exponent varies non-smoothly
as the parameter $E$ is taken from allowed and to forbidden values.
This in principal, could provide a new method to calculate the allowed
energy spectrum of the eigenvalue problem.

As seen in Fig. 2, below criticality,
the allowed energies give rise to quasiperiodic orbits (invariant curves)
with zero Lyapunov exponent while the states in the forbidden
regime are characterized by negative Lyapunov exponents. In the
superctical regime, the Lyapunov exponent is always negative. 
At criticality, where the spectrum is conjectured to be singular
continuous \cite{Mathieu}, the plot
clearly illustrates the fact that the allowed energies form a Cantor
set of zero measure.
Another interesting aspect of the driven map is that 
the inverse localization length of the Harper equation gives
the absolute value of the Lyapunov exponent. Since the
localization length of the system is exactly known, this is an example
of a system whose Lyapunov exponent is known exactly.

Finally, comparing Figs. 2 (a) and (b),
we see a very interesting manifestation of the
self-duality \cite{Mathieu} of the Harper equation
which can be written as
\begin{equation}
\gamma ( \lambda, E ) = -ln(\lambda) + \gamma(1/ \lambda, E/ \lambda ) .
\end{equation}

Fig. 3 shows the attractor of the mapping (\ref{map}) at and slightly
away from criticality. The self-similarity of the Harper wave function at
critically implies that the attractor is fractal with structure at all
scales. Interesting aspect of this attractor is the fact that it
has a zero Lyapunov exponent. Therefore, the map exhibits a new
type of SNA where the two nearby trajectories converge on the attractor, not
exponentially but with a power law. In the supercritical regime, the
attractor is associated with negative Lyapunov exponent which appears
to smooth out the fragmented structure of the attractor.
However, as we will argue later in this section, the attractor is
in fact strange throughout the supercritical regime.

Another intriguing result emerging from this mapping is that
in the localized regime, the critical phase $\phi_c$ of
the potential leading to localization around the site
 $i=0$ can be found as a ``homoclinic''  point where
the forward and backward attractors of the map intersect (see Fig. 4).
It is interesting to note that independently of the value of $\lambda > 1$,
a homoclinic point is always found at $\phi_c =1/4$. 
In the strong coupling limit,
the homoclinic points coincide with the divergences seen in
the attractor. This point of view
brings out the significance of the critical phase factor in the Harper
equation which has not been understood in the past.
Unlike the supercritical regime, at the critical point,
the forward and backward attractors seem to overlay.

In the localized regime,
the attractor is strange but nonchaotic.
The appearance of the fractal structure can be 
easily seen coming down from the strong coupling
limit $\lambda \to \infty$. Here we give a rather crude analysis of this
problem showing just the main ideas of the argument.

Let us first rewrite the map in terms of the
variables $x$ and $\theta$ assuming $E=0$:
\begin{eqnarray}
x'  = {-1 \over x + 2\lambda \cos(2\pi \theta)} \\
\theta ' =\theta + \sigma \;\;mod \;\; 1.\label{map2v}
\end{eqnarray}
Note that the map remains invariant under the transformation
$x \to -x$, $\theta \to \theta + 1/2$ so we can expect the
attractor to have this symmetry as well. Therefore, in the following
we assume implicitly that $\theta $ is considered mod $1/2$.

To the first order in $1/\lambda$,
the attractor, whose invariant measure is always uniform in $\theta$,
can be written as 
\begin{equation} 
x (\theta ) = {-1 \over  2\lambda \cos[2\pi (\theta-\sigma )]} . \label{0appr}
\end{equation}
However, the above form suggests a singularity appearing 
asymptotically as $\lambda \to \infty$ at
$\theta_1 = 1/4 +\sigma $.
This first order singularity gives rise to higher order
singularities which are absent in the above approximation for
the attractor.
In order to see the appearance of the second order singularity,
note that according to Eq. (\ref{0appr}),
$|x(\theta )|$ takes values in the interval $[1/(2\lambda ) ,\infty)$
when $\theta$ is varied in $[0,1/2)$. Thus, it is possible to find $\theta_2^*$
such that
$x(\theta_2^* -\sigma) +2\lambda \cos(2\pi \theta_2^* ) =0$ which gives
rise to a singularity around
$\theta_2 = \theta_2^* +\sigma $. Moreover, asymptotically
$\theta_2^* \to \theta_1 +\sigma$ because the singularity around
$\theta_1$ becomes very sharp and the attractor is mostly (as far as the
invariant measure is concerned) close to zero away  
from $\theta_1$ as $\lambda \to \infty$.
In other words, we have generated a new singularity
close to $\theta_2 = 1/4 +2\sigma $. Looking
at the explicit (asymptotic) equation for $\theta_2^*$,
\begin{equation}
4\lambda^2 \cos(2\pi \theta_2^*)\cos[2\pi (\theta_2^*-\sigma )] =1,
\end{equation}
shows that the second order singularity is even sharper than
the first order one. Now the same argument can be repeated
to generate a third order singularity from the second order one
and so on. Each new singularity is sharper than the previous one.
Because the measure is uniform in $\theta$, the high order
singularities become invisible for finite number of iterations of the map.
Asymptotically with $\lambda \to \infty$,
 the singularities generated in this way appear
at $\theta_m = 1/4 + m\sigma $ $(m=1,2,...)$.
For finite $\lambda$, their locations move a little bit but
we still expect them to be dense in $\theta$.
Their existence causes the attractor to be nowhere
differentiable and to have the characteristics of a SNA \cite{GOPY}.

In summary, the above reasoning provides a very strong argument for the
existence of SNA in quasiperiodically driven maps given by Eq.
(\ref{map}).
 
\section{Discussion}

In this paper, we have shown an interesting relationship between
the supercritical Harper equation
and the critical dissipative standard map: 
the fluctuations
in the localized eigenstates of the Harper equation
are related to the tangent orbit of the
standard map where the parameter $1/\lambda$ plays the role of 
the dissipation parameter $b$. In particular, the strong coupling limit
of the Harper equation 
is analogous to the strong dissipation (or circle map) limit of the standard
map. Inspite of these similarities, the two problems
are quantitatively different due to the fact that they are
characterized by different universal
numbers $\zeta$. The roots for these two quantitatively different universality
classes is tied to the fact that the dynamics governing the nonlinear
potential in the Harper equation is a pure rotation while the dynamics
underlying the  critical dissipative standard map is highly
nontrivial.

The same kind of reasoning can be applied also to
another related class of problems studied by us recently
\cite{KSphon}, namely the phonon modes
of the Frenkel-Kontorova model where the locations
of the particles are described by the
area-preseving standard map. Although this problem bears a close
resemblance to the Harper equation,
the supercritical regime in the Frenkel-Kontorova model
has very different characteristics.
Below criticality, the
potential is derived from the dynamics analytically conjugate to the
pure rotation in both models. However, beyond criticality, 
the Harper potential is still given by the pure rotation whereas
the phonon potential is obtained from cantorus dynamics which 
cannot be mapped smoothly to the rotation. 
In the Harper equation, the ``supercritical'' regime is described
by exponentially localized eigenfunctions with 
universal self-similar fluctuations characterized by a unique strong
coupling fixed point \cite{KSloc}.
In contrast, the phonon eigenmodes defy localization
and remain critical with scaling characterized by a line of
renormalization limit cycles.
In addition, there exist an infinite sequence of
parameter values in the supercritical regime
where the renormalization limit cycle degenerates into a trivial 
fixed point. 
A very intriguing characteristics of these parameter values is the fact that
the corresponding phonon eigenfunction is represented by an infinite
series of step-functions.

Our studies suggest that the solutions of eigenvalue problems
where the potential is determined by
sequences with complicated dynamics can be extremely rich. By studying
the Harper equation 
and the phonons in the Frenkel-Kontorova model, we may have barely 
scratched the surface of the wealth of novel phenomena exhibited
by these types of systems. For example, in the generalized versions of
the Harper equations one encounters fat critical regimes with  
ergodic renormalization dynamics in the strong coupling limit
\cite{KS,KSloc}.

The idea of relating an almost periodic eigenvalue problem 
to a quasiperiodically driven map 
has opened new possibilities for
studying SNA. In this paper, we have shown that
the existence of localization in the
eigenvalue problem is associated with the appearance of
homoclinic points in the corresponding map.
Furthermore, the critical state at the onset to localization
is found to correspond to a new type of strange attractor
with zero Lyapunov exponent. In future, we also hope
to answer the question about the class of models whose whole
supercritical phase is described by a unique strong coupling limit
of the theory. In particular, it would be interesting to know whether
quasiperiodically  driven maps studied earlier exhibit this type
of universality or not. 

The mapping from linear-difference equations to
driven maps is of continuing interest to us.
In future, we would like to apply this technique to a variety of other problems
that can be written in the linear-difference form.
In particular, the one to two-hole transition in the extended standard
map \cite{BM}, which corresponds to
nonexponentially localized phonons \cite{KSphon},
may provide new dynamics in the forced map.
Fishman et al. \cite{FGP} have shown that the quantum dynamics of the kicked
rotor can be mapped to the tight binding model of the above form.
The integrable rotor with quasiperiodic potential
has been shown to exhibit Harper-type universality.
However, according to our preliminary study on the nonintegrable rotor with
pseudorandom potential, the localized phase is not characterized by a unique
strong coupling fixed point in this model.

The research of IIS is supported by a grant from National Science
Foundation DMR~093296.
JAK would like to thank the organizers of the conference for
their kind invitation.

\begin{figure}
\caption{The renormalization fixed point for the fluctuations of
the supercritical Harper equation obtained by solving the expansion
of $1/e(x)$ upto the order $x^{23}$ by the Newton method.
The fact that there has to be a singularity (or zero point
for $1/e$) can be seen easily from the fixed point equation.
The horizontal line shows the value of the universal $\zeta_H$.}
\label{fig1}
\end{figure}

\begin{figure}
\caption{Lyapunov exponent $\gamma$
vs the parameter E for the one-dimensional map
derived from the Harper equation. The figure obtained with
$\lambda=0.5$ (a) is equivalent to the figure with $\lambda=2$ (b)
after simple scaling of $\gamma$ and $E$ which illustrates the self-duality
of the Harper equation. Fig. (c) shows the corresponding results at
$\lambda=1$. The allowed values of the energy are characterized by an attractor
with zero Lyapunov exponent.}
\label{fig2}
\end{figure}

\begin{figure}
\caption{The attractor of the one-dimensional map ($E=0$) at
the critical point $\lambda=1$ (a) and at $\lambda=1.05$ (b).}
\label{fig3}
\end{figure}

\begin{figure}
\caption{The forward (lighter dots) and backward (darker dots)
 attractors of the
map and the existence of homoclinic points for $\lambda =2$. As
$\lambda$ approaches $\infty$, the divergences coincide with
the homoclinic points.}
\label{fig4}
\end{figure}

\end{document}